\pdfoutput=1

\documentclass[twoside,twocolumn,9pt]{article}
\usepackage{extsizes}
\usepackage[super,sort&compress,comma]{natbib} 
\usepackage[version=3]{mhchem}
\usepackage[left=1.5cm, right=1.5cm, top=1.785cm, bottom=2.0cm]{geometry}
\usepackage{balance}
\usepackage{times,mathptmx}
\usepackage{sectsty}
\usepackage{graphicx} 
\usepackage{subcaption}
\usepackage{amsmath,bm}
\usepackage{lastpage}
\usepackage[format=plain,justification=justified,singlelinecheck=false,font={stretch=1.125,small,sf},labelfont=bf,labelsep=space]{caption}
\usepackage{float}
\usepackage{fancyhdr}
\usepackage{fnpos}
\usepackage[english]{babel}
\usepackage{array}
\usepackage{droidsans}
\usepackage{charter}
\usepackage[T1]{fontenc}
\usepackage[usenames,dvipsnames]{xcolor}
\usepackage{setspace}
\usepackage[compact]{titlesec}
\usepackage{dblfloatfix}
\usepackage{color,soul}
\usepackage{comment}

\usepackage{epstopdf}

\definecolor{cream}{RGB}{222,217,201}

\begin{document}
\setlength{\parskip}{0pt}
\pagestyle{fancy}
\thispagestyle{plain}
\fancypagestyle{plain}{

\renewcommand{\headrulewidth}{0pt}
}

\makeFNbottom
\makeatletter
\renewcommand\LARGE{\@setfontsize\LARGE{15pt}{17}}
\renewcommand\Large{\@setfontsize\Large{12pt}{14}}
\renewcommand\large{\@setfontsize\large{10pt}{12}}
\renewcommand\footnotesize{\@setfontsize\footnotesize{7pt}{10}}
\makeatother

\renewcommand{\thefootnote}{\fnsymbol{footnote}}
\renewcommand\footnoterule{\vspace*{1pt}%
\color{cream}\hrule width 3.5in height 0.4pt \color{black}\vspace*{5pt}} 
\setcounter{secnumdepth}{5}

\makeatletter 
\renewcommand\@biblabel[1]{#1}            
\renewcommand\@makefntext[1]%
{\noindent\makebox[0pt][r]{\@thefnmark\,}#1}
\makeatother 
\renewcommand{\figurename}{\small{Fig.}~}
\sectionfont{\sffamily\Large}
\subsectionfont{\normalsize}
\subsubsectionfont{\bf}
\setstretch{1.125} 
\setlength{\skip\footins}{0.8cm}
\setlength{\footnotesep}{0.25cm}
\setlength{\jot}{10pt}
\titlespacing*{\section}{0pt}{4pt}{4pt}
\titlespacing*{\subsection}{0pt}{15pt}{1pt}

\renewcommand{\headrulewidth}{0pt} 
\renewcommand{\footrulewidth}{0pt}
\setlength{\arrayrulewidth}{1pt}
\setlength{\columnsep}{6.5mm}
\setlength\bibsep{1pt}

\makeatletter 
\newlength{\figrulesep} 
\setlength{\figrulesep}{0.5\textfloatsep} 

\newcommand{\topfigrule}{\vspace*{-1pt}%
\noindent{\color{cream}\rule[-\figrulesep]{\columnwidth}{1.5pt}} }

\newcommand{\botfigrule}{\vspace*{-2pt}%
\noindent{\color{cream}\rule[\figrulesep]{\columnwidth}{1.5pt}} }

\newcommand{\dblfigrule}{\vspace*{-1pt}%
\noindent{\color{cream}\rule[-\figrulesep]{\textwidth}{1.5pt}} }

\makeatother

\twocolumn[
  \begin{@twocolumnfalse}
\vspace{3cm}
\sffamily
\begin{tabular}{m{4.5cm} p{13.5cm} }

& \noindent\LARGE{\textbf{Excited-State Solvation Structure of Transition Metal Complexes from Molecular Dynamics Simulations and Assessment of Partial Atomic Charge Methods}} \\

\vspace{0.3cm} & \vspace{0.3cm} \\

 & \noindent\large{Mostafa Abedi,$^{a}$ Gianluca Levi,$\ddag ^{a}$ Diana B. Zederkof,$^b$ Niels E. Henriksen,$^{a}$  M\'{a}ty\'{a}s P\'{a}pai,$^{ac}$ and Klaus B. M{\o}ller$^{\ast^a}$} \\

& \noindent\normalsize{In this work, we investigate the excited-state solute and solvation structure of $\mathrm{[Ru(bpy)_3]^{2+}}$, $\mathrm{[Fe(bpy)_3]^{2+}}$, $\mathrm{[Fe(bmip)_2]^{2+}}$ and $\mathrm{[Cu(phen)_2]^{+}}$ (bpy=2,2'-pyridine; bmip=2,6-bis(3-methyl-imidazole-1-ylidine)-pyridine; phen=1,10-phenanthroline) transition metal complexes (TMCs) in terms of solute-solvent radial distribution functions (RDFs) and evaluate the performance of some of the most popular partial atomic charge (PAC) methods for obtaining these RDFs by molecular dynamics (MD) simulations. To this end, we compare classical MD of a frozen solute in water and acetonitrile (ACN) with quantum mechanics/molecular mechanics Born-Oppenheimer molecular dynamics (QM/MM BOMD) simulations. The calculated RDFs show that the choice of a suitable PAC method is dependent on the coordination number of the metal, denticity of the ligands, and type of solvent. It is found that this selection is less sensitive for water than ACN. Furthermore, a careful choice of the PAC method should be considered for TMCs that exhibit a free direct coordination site, such as $\mathrm{[Cu(phen)_2]^{+}}$. 
The results of this work show that fast classical MD simulations with ChelpG/RESP or CM5 PACs can produce RDFs close to those obtained by QM/MM MD and thus, provide reliable solvation structures of TMCs to be used, e.g. in the analysis of scattering data.} \\

\end{tabular}

 \end{@twocolumnfalse} \vspace{0.6cm}

  ]

\renewcommand*\rmdefault{bch}\normalfont\upshape
\rmfamily
\section*{}
\vspace{-1cm}


\footnotetext{\textit{$^{a}$Department of Chemistry, Technical University of Denmark, 2800 Kongens Lyngby, Denmark; E-mail: kbmo@kemi.dtu.dk \newline 
${\ddag}$ Present address: Faculty of Physical Sciences, University of Iceland, 107 Reykjav\'{i}k, Iceland \newline
$^{b}$Department of Physics, Technical University of Denmark, 2800 Kongens Lyngby, Denmark \newline
$^{c}$Wigner Research Center for Physics, Hungarian Academy of Sciences, P.O. Box 49, H-1525 Budapest, Hungary}}



\section{Introduction}

Transition metal complexes (TMCs) have been used broadly in solar energy conversion and photocatalysis applications due to their excited-state photophysical and photochemical properties.\cite{Balzani2007} The advent and development of ultrafast spectroscopy in recent years has made it feasible to study and unravel mechanisms of ultrafast excited-state dynamics of TMCs in solution.\cite{doi:10.1021/ar500358q,doi:10.1021/ar030111d,Iwamura2015UltrafastComplexes} The experiments reveal that excited-state photophysical and photochemical properties can be strongly affected by the molecular environment in solution leading to significant changes in relaxation rates and products.\cite{Penfold2012,Nahhas2010,Dohn2016,Kjaer2018,Chagas2014SolventRuNH35Pyrazine2+,C5RA25670D,doi:10.1021/jp2022715,AGENA201760} Therefore, in order to get a deep understanding of such processes, a detailed insight into the effect and role of the solvent is essential. By utilizing  time-resolved X-ray diffuse scattering (TRXDS) and extended X-ray absorption fine structure (EXAFS) spectroscopy, one can obtain information about the solvent structure.\cite{Haldrup2016ObservingScattering,doi:10.1021/jp306917x,PhysRevLett.117.013002,VanDriel2016,Canton2015} However, the complicated ultrafast excited-state dynamics of TMCs and, in particular, the nearest surrounding solvation make the interpretation of the observed experimental features difficult. In this regard, computational chemistry tools play an essential role for comprehensive interpretation and understanding. 

There are two commonly used theoretical approaches for considering solvent effects: methods that treat the solvent explicitly and implicit models. In the implicit solvent treatment, also known as continuum solvation models, \cite{doi:10.1021/cr9904009,doi:10.1021/cr960149m,doi:10.1002/qua.25725} the solvent molecules are approximated by a homogeneous medium. The polarizable continuum model (PCM)\cite{doi:10.1021/cr9904009} is a well-known implicit solvent model. Implicit models do not provide information about the structure of the solvent. Solvent structure can only be simulated using explicit solvent methods, in which the interaction between all solute-solvent pairs is explicitly considered. The explicit solvent treatment is usually employed in classical molecular dynamics (MD), \textit{ab initio} MD (AIMD) or hybrid quantum mechanics/molecular mechanics (QM/MM) MD simulation approaches. In classical MD simulations, the forces are obtained from predetermined molecular mechanics (MM) force fields, while in AIMD an electronic structure method is used for calculating the forces on-the-fly; in QM/MM MD, a hybrid scheme of these two approaches is applied. Classical MD simulations are among the most popular methods for studying chemical processes of medium- to large-size systems in condensed phases. One of the main challenges in classical MD is the specification of suitable empirical models for the forces between the atoms. These force fields are parametrized by fitting to experimental data or high-level \textit{ab initio} calculations. A major limitation of this method is that the model is not transferable to any type of reaction or chemical process and often needs to be re-parameterized. On the other hand, classical MD simulations are fast and easy to handle. 
AIMD addresses the limitation of force fields in MD simulations. In this method, the forces are calculated "on the fly", i.e. during the MD propagation,  from electronic structure calculations, typically using density functional theory (DFT). Because the electronic structure calculation is performed at every time step of the simulation, the AIMD method is computationally very demanding. The computational cost can be reduced by density functional tight binding (DFTB),\cite{PhysRevB.58.7260} which is much faster than DFT but less accurate\cite{doi:10.1021/acs.jpca.7b10664} or by QM/MM MD, in which the most important part of the system 
is described by a suitable (high-level) quantum chemistry method and the rest by molecular mechanics using a force field\cite{WARSHEL1976227,doi:10.1002/jcc.540110605,Groenhof2013,doi:10.1002/anie.200802019}. 

Solvation structure can be obtained from explicit solvent methods through the evaluation of solute-solvent radial distribution functions (RDFs), which can be used for calculations of XDS\cite{0953-4075-48-24-244010} and EXAFS\cite{doi:10.1063/1.466581} signals. Classical MD simulations have been extensively applied for calculating RDFs and gaining information about the solvation structure around TMCs in both ground and excited states.\cite{doi:10.1063/1.4939898,C3CP44465A,doi:10.1021/acs.jpcb.5b10980,C5CP06406F,doi:10.1021/jp2022715,doi:10.1021/jp403751m,doi:10.1021/acs.jpca.5b03842,doi:10.1021/jp900147r}
However, standard available force fields are particularly developed for ground-state (GS) MD simulations. This rises a serious problem when performing MD simulations in the excited state: the force fields essentially are required to be re-parametrized. The pairwise electrostatic interactions between solute and solvent atoms, which rely on the choice of partial atomic charges (PACs), play a key role in the determination of solvent configurations in MD simulations. In this work, we explore the idea of using PACs of the excited state of the solute from DFT calculations in MD simulations while keeping the GS van der Waals (vdW) parameters, to develop an approximate excited-state force field. For polar solvents like water and ACN this is justified by the fact that the contribution of vdW terms (non-electrostatic non-bonded interactions) is significantly smaller than those of electrostatic interactions in the potential energy of the system.\cite{doi:10.1146/annurev-physchem-050317-021013} Therefore, using the GS vdW parameters for the excited-state simulations should not result in a notable error. 

In the present work, we perform classical MD simulations and assess the performance of several of the most popular PAC methods (see Theoretical Methods part) in the description of the solvation structure of four  prototypical polypyridine TMCs including two \textit{tris}-bidentate TMCs $\mathrm{[Ru(bpy)_3]^{2+}}$ and $\mathrm{[Fe(bpy)_3]^{2+}}$, a \textit{bis}-tridentate TMC $\mathrm{[Fe(bmip)_2]^{2+}}$ and a \textit{bis}-bidentate TMC $\mathrm{[Cu(phen)_2]^{+}}$
(bpy=2,2'-pyridine; bmip=2,6-bis(3-methyl-imidazole-1-ylidine)-pyridine; phen=1,10-phenanthroline) (see Fig. 1). The excited-state dynamics of these TMCs have been extensively investigated.\cite{doi:10.1021/jp2022715,CHEM:CHEM201000184,doi:10.1021/acs.jpclett.7b01479,doi:10.1021/jz100548m,Aubock2015Sub-50-fsFebpy32+,C5CP06406F,C3CC43833C,Papai2016High-EfficiencyDynamics,Iwamura2015UltrafastComplexes,doi:10.1021/acs.jpca.5b03842} These TMCs represent a comprehensive set exhibiting a range of possibilities for solvent molecules to approach the metallic center depending on the coordination number of the metal and denticity of the ligands. We assess the performance of the various PAC methods by contrasting the RDFs simulated by classical MD with frozen solute to those obtained by QM/MM Born-Oppenheimer molecular dynamics (BOMD) simulations of a non-rigid solute carried out in the present work or taken from the literature (QM/MM MD and AIMD). In this work, we seek suitable PAC methods, which enable us to perform classical MD simulations (with frozen solute) without need of force field reparameterization to provide reliable RDFs. These results can be used to complement and assist experimental determinations. 
The QM/MM BOMD and classical MD simulations are performed in water for $\mathrm{[Ru(bpy)_3]^{2+}}$ and $\mathrm{[Fe(bpy)_3]^{2+}}$ and in acetonitrile (ACN) for $\mathrm{[Fe(bmip)_2]^{2+}}$ and $\mathrm{[Cu(phen)_2]^{+}}$; these two solvents are the most popular ones in experimental studies of such TMCs. 
    
\begin{figure}[h]
    \centering
    \begin{subfigure}[b]{0.15\textwidth}
        \includegraphics[height=2.5cm]{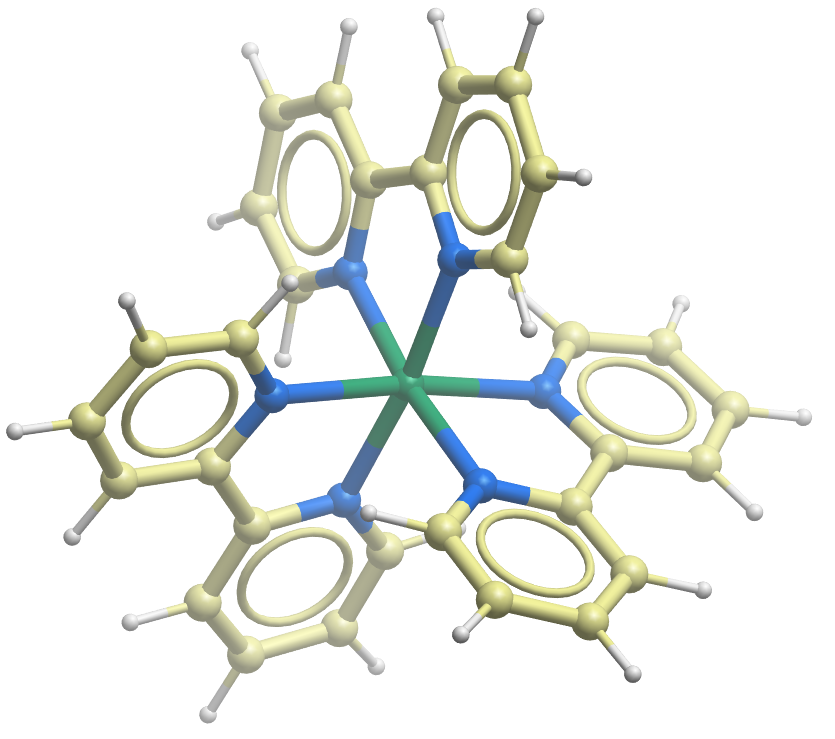}
        \caption{$\mathrm{[Ru/Fe(bpy)_3]^{2+}}$}
    \end{subfigure}
    \begin{subfigure}[b]{0.15\textwidth}
        \includegraphics[height=2.5cm]{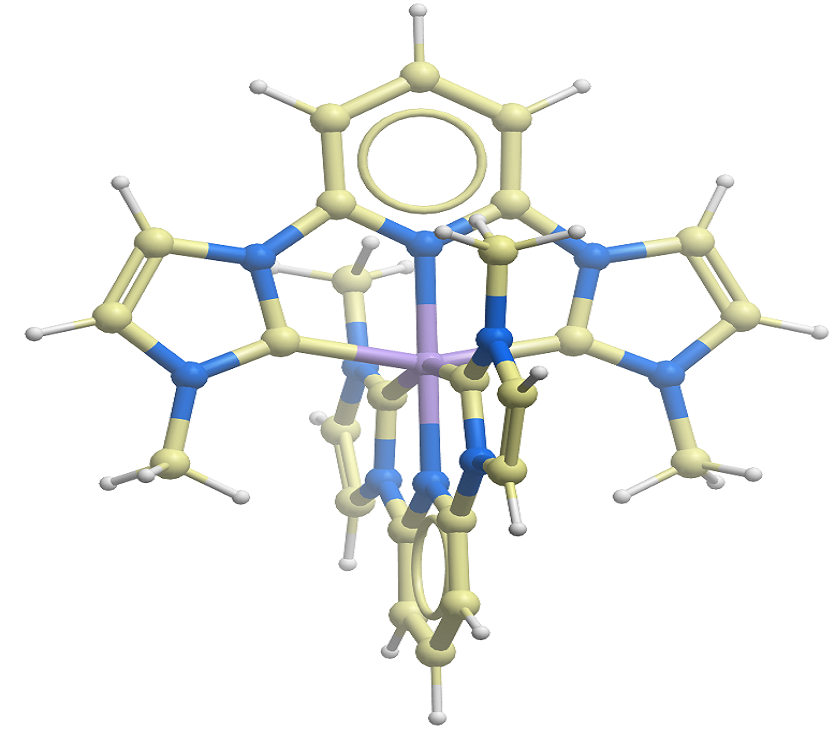}
        \caption{$\mathrm{[Fe(bmip)_2]^{2+}}$}
    \end{subfigure}
    \begin{subfigure}[b]{0.15\textwidth}
        \includegraphics[height=2.5cm]{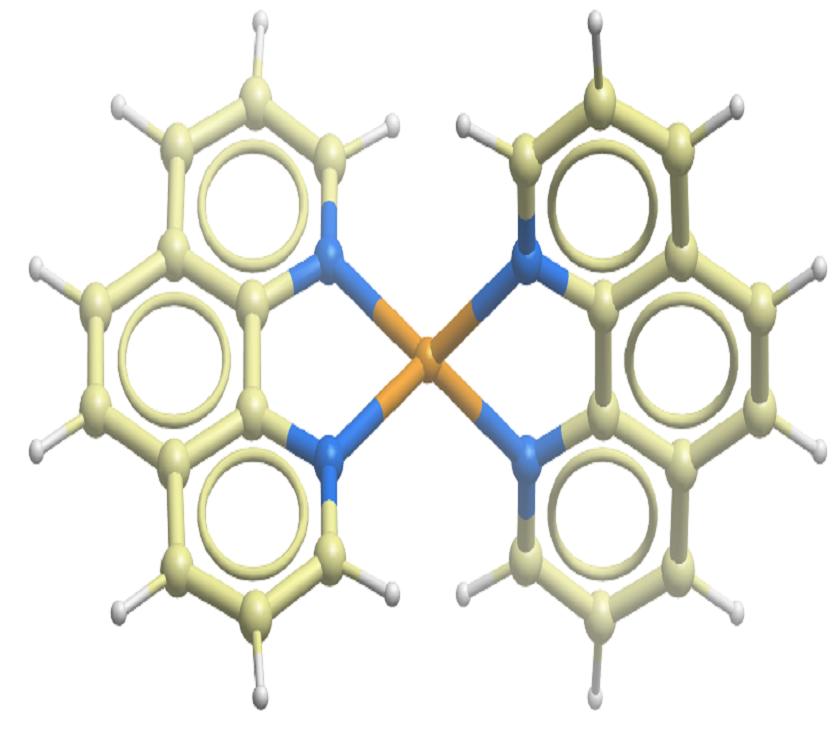}
        \caption{$\mathrm{[Cu(phen)_2]^{+}}$}
    \end{subfigure}
    \caption{Schematic molecular structures of the investigated TMCs. Color codes: Nitrogen--blue; Carbon--yellow; Hydrogen--white; Ruthenium--green; Iron--violet; Copper--orange.}
\end{figure}

\section{Theoretical Methods}

\subsection{Partial Atomic Charges}

As PACs are not quantum mechanical observables, many different methods have emerged to calculate them. In the present work, seven common PAC methods have been chosen: Mulliken population analysis (MPA)\cite{doi:10.1063/1.1740588}, natural population analysis (NPA)\cite{doi:10.1063/1.449486}, charges from electrostatic potentials using a grid based method (ChelpG)\cite{JCC:JCC540110311}, restrained electrostatic potential (RESP)\cite{doi:10.1021/j100142a004}, atoms in molecules (AIM)\cite{doi:10.1021/cr00005a013}, Hirshfeld\cite{Hirshfeld1977} and charge model 5 (CM5)\cite{doi:10.1021/ct200866d}.

MPA and NPA are methods based on partitioning the molecular electronic wave function. MPA, due to its simplicity, is the most straightforward method for assigning PACs and almost all quantum chemistry programs provide it as default population analysis. However, this method suffers from basis-set dependency and lack of convergence of atomic charges with increasing basis-set size. 
NPA was developed by Reed \textit{et al.}\cite{doi:10.1063/1.449486} as an alternative to overcome the problems with MPA. This method works based on natural atomic orbitals on each atomic center which are orthogonal and less sensitive to the basis set. The NPA method is usually recommended for characterization of the electron distribution in systems that have high ionic character.\cite{doi:10.1063/1.449486}

ChelpG and RESP, in which PACs are derived through a fitting procedure to reproduce the molecular electrostatic potential (ESP), a real physical observable, are among the most popular methods for assigning atomic charges. The ESP at a given point $\textit{i}$ is computed by eqn (1):
\begin{equation}
 \Phi_{ESP}(\bm{r}_i) = \sum_{\alpha}^{M} \frac{Z_\alpha}{|\bm{R}_\alpha - \bm{r}_i|} - \sum_{j}^{L} \frac{\rho(\bm{r}_j)}{|\bm{r}_j-\bm{r}_i|} V_p
\end{equation}
where \textit{Z}$_\alpha$ and $\bm{R}_\alpha$ are respectively the charge and position of nucleus $\alpha$ and $M$ is the total number of nuclei. The electron density of the molecule at point $j$ is denoted by ${\rho(\bm r}_j)$, where $\bm{r}_j$ is the grid point coordinate. $V_p$ is the volume per grid point and \textit{L} is the total number of grid points. The first term in eqn (1) is straightforward to calculate. But for the molecular electron density, quantum chemistry calculations are required. Atomic charges $Q_\alpha$ are obtained by least squares fitting of the molecular ESP.
The best fit is achieved by minimization of an error function, $F_{error}^{ESP}$, (eqn (2)) so that the ESP predicted by the $Q_\alpha$ is as close as possible to $\Phi_{ESP}$.
\begin{equation}
 F_{error}^{ESP} = \sum_{i}^{N} \Big[\Phi_{ESP}(\bm{r}_i) - \sum_{\alpha}^{M}\frac{Q_\alpha}{|\bm{R}_\alpha - \bm{r}_i|} \Big]^2
\end{equation}
Here, $N$ is the total number of ESP points. The $Q_\alpha$ ($\alpha$=1, ..., $M$) can be found by solving eqn (3):
\begin{equation}
\frac{\partial F_{error}^{ESP}}{\partial Q_\alpha} = - \sum_{i}^{N} \frac{2}{{|\bm{R}_\alpha - \bm{r}_i|}}  \Big[\Phi_{ESP}(\bm{r}_i) - \sum_{\alpha}^{M}\frac{Q_\alpha}{|\bm{R}_\alpha - \bm{r}_i|} \Big] = 0
\end{equation}
In the ChelpG method, a cubic box is designed and the molecular ESP points are generated between 0-2.8 {\AA} from the vdW surface of the molecule. A well-known issue that affects ChelpG is the poor prediction of the atomic charges of deeply buried atoms, such as metals in TMCs. This is because during the fitting procedure the molecular ESP points are far from the buried atoms. This problem is addressed by the RESP method by utilizing a penalty function in eqn (2), which enables us to introduce target charges and the possibility to fix them during the fitting. Moreover, this method ensures that atoms with the same chemical environment possess identical partial charges. Here, we use a hyperbolic penalty function. Bayly \textit{et al.}\cite{doi:10.1021/j100142a004} have found that a hyperbolic restraint function determines charges better than a quadratic function. Eqn (4) shows the modified error function for the calculation of the RESP charges:
\begin{equation}
\begin{split}
 F_{error}^{RESP} = \sum_{i}^{N}  \Big[\Phi_{ESP}(\bm{r}_i) - \sum_{\alpha}^{M}\frac{Q_\alpha}{|\bm{R}_\alpha - \bm{r}_i|} \Big]^2 \\ 
 + \beta \sum_{\alpha}^{M} \Big[\sqrt{(Q_{0 \alpha} - Q_\alpha)^2 + b^2} - b \Big]
\end{split}
\end{equation}
Here, $\beta$ is a quantity for setting the strength of the restraint, $Q_{0 \alpha}$ is the target charge and  \textit{b} is the tightness of the hyperbola around its minimum. The RESP charges can be obtained by solving eqn (5):

\begin{equation}
\begin{split}
\frac{\partial F_{error}^{RESP}}{\partial Q_\alpha} =  -\sum_{i}^{N}  \frac{2}{{|\bm{R}_\alpha - \bm{r}_i|}}  \Big[\Phi_{ESP}(\bm{r}_i) - \sum_{\alpha}^{M}\frac{Q_\alpha}{|\bm{R}_\alpha - \bm{r}_i|} \Big] \\ 
- \beta \frac{(Q_{0 \alpha} - Q_\alpha)}{{\sqrt{(Q_{0 \alpha} - Q_\alpha)^2 + b^2}}}= 0
\end{split}
\end{equation} 

The basis of the AIM and Hirshfeld methods is to partition the electron density into atomic domains. In the AIM method, topological analysis of the electron density is used to find the electron density maxima (which often occur at the nuclei) and minima. The atomic domains (also known as Bader regions) are obtained by following the density gradients. The border between regions, which is called the zero flux surface, is placed where the density gradient is zero. The partial charges are then obtained by integration of the electron density in each atomic domain. The Hirshfeld method is similar to AIM except that the atomic domains are defined based on a weight factor, which is the ratio of the electron densities of isolated atoms and the density constructed from a sum of atomic densities (the so-called promolecular density). The main disadvantages of the AIM method are its computational cost and the overestimation of partial charges for polar bonds\cite{doi:10.1002/9780470125823.ch3}, similarly to the NPA method\cite{SANNIGRAHI199373}, while the Hirshfeld method frequently underestimates these charges\cite{Davidson1992}.  
Finally, CM5 is a parametrized method that uses gas-phase Hirshfeld charges as input and derives PACs to reproduce the molecular dipole moment. The charges derived by dividing the electron density are less sensitive to the basis set size and usually yield more reasonable PACs for the buried atoms.            

 \begin{figure*}[h]
    \centering
    \begin{subfigure}[b]{0.8\textwidth}
        \includegraphics[width=\textwidth]{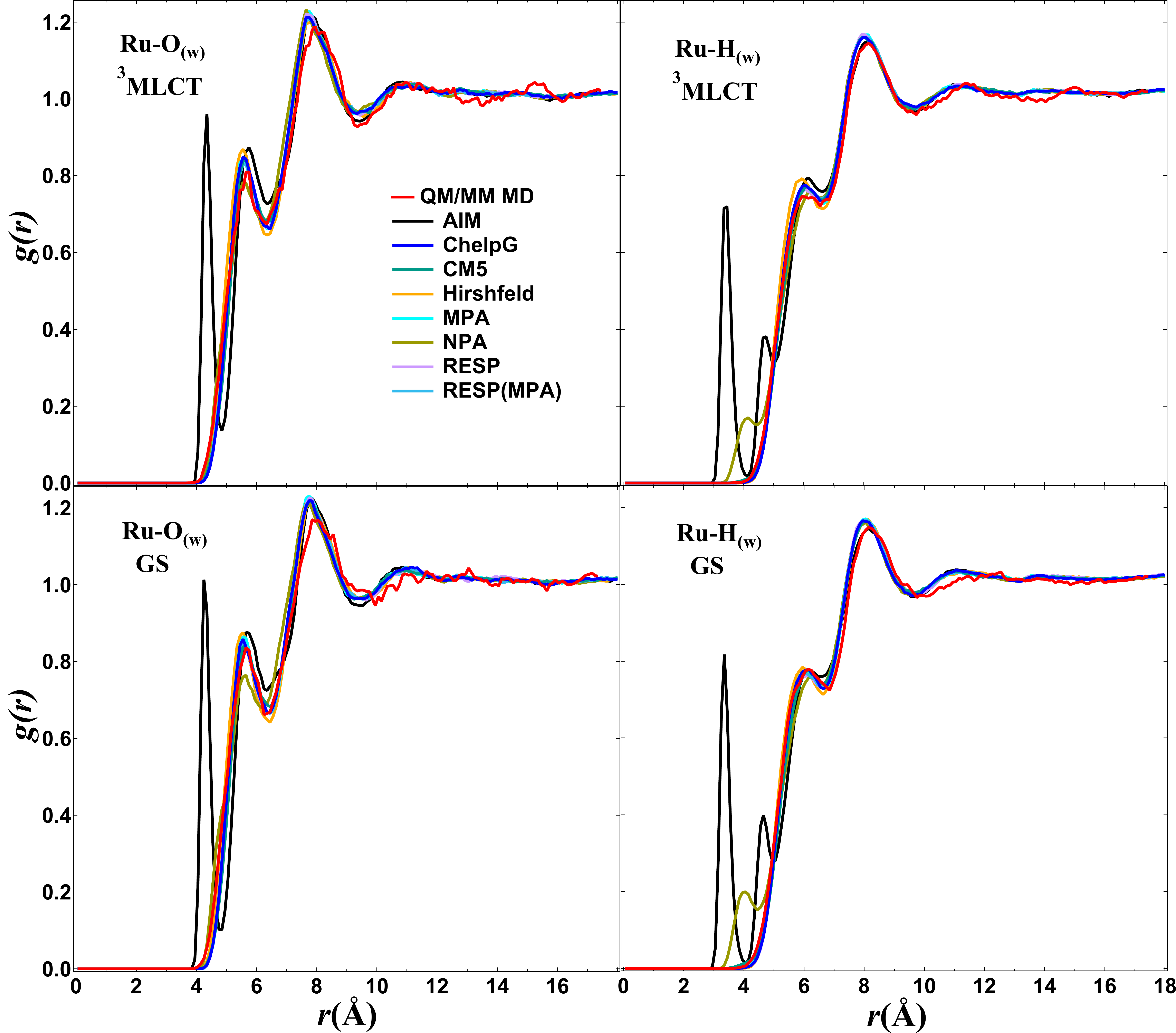}
    \end{subfigure}
   \caption{ The RDFs, $g(r)$, of $\mathrm{[Ru(bpy)_3]^{2+}}$ in water for the Ru-$\mathrm{O_w}$ and Ru-$\mathrm{H_w}$ pairs obtained from the classical MD and QM/MM MD (red lines) simulations in the GS and $^3$MLCT state.}
\end{figure*} 

\subsection{Computational Details}

In this section, we provide the computational details for the methods used in this work. Section 2.2.1 covers the geometry optimizations and PAC calculations of the chosen TMCs using density functional theory (DFT) in gas phase. In section 2.2.2, we discuss the classical MD simulations utilizing the optimized structures and PACs obtained from DFT calculations. Finally, a detailed description of the QM/MM MD simulations is given in section 2.2.3.   

\subsubsection{Electronic Structure Calculations: Optimizations and PAC Calculations}
The structures of the four selected TMCs were optimized using DFT with the B3LYP* hybrid exchange-correlation functional\cite{Reiher2001,doi:10.1021/ic025891l} in combination with a triple zeta valence quality basis set augmented by polarization functions (Def2TZVP)\cite{B508541A}.  The B3LYP* functional has been benchmarked for the structural and energetic characteristics of TMCs against high-level quantum chemical methods and experimental results and shown reliable performance.\cite{doi:10.1021/ct300932n,C7SC02815F,LawsonDaku2005AssessmentComplex,Reiher2002TheoreticalFunctionals,Paulsen2013ProgressComplexes} $\mathrm{D_{2d}}$ and $\mathrm{C_2}$ symmetries are used for $\mathrm{[Fe(bmip)_2]^{2+}}$ in its ground and excited states, respectively, and $\mathrm{C_1}$ for the other TMCs (geometry optimizations and PAC calculations for classical MD simulations).\par 
For the GS calculations, the total spin angular momentum quantum number was set to zero ($S$=0) for all structures while unrestricted open-shell calculations were performed for the  low-lying triplet metal-to-ligand charge-transfer ($^3$MLCT) state (for $\mathrm{[Ru(bpy)_3]^{2+}}$ and $\mathrm{[Cu(phen)_2]^{+}}$) and low-lying quintet metal-centered, $^5$MC (high-spin; HS) state (for $\mathrm{[Fe(bpy)_3]^{2+}}$), applying $S$=1 and $S$=2, respectively. 
It should be noted that the very high density of low-lying electronic states in $\mathrm{[Fe(bmip)_2]^{2+}}$ leads to several conical intersections between the $^3$MLCT and $^3$MC states\cite{doi:10.1021/jz500829w,Papai2016High-EfficiencyDynamics} which prevents us from performing state-specific QM/MM BOMD simulations in the excited state. Therefore, for $\mathrm{[Fe(bmip)_2]^{2+}}$ we only compare the GS RDFs obtained from the classical MD simulations with the QM/MM BOMD ones. The Cartesian coordinates of the GS DFT optimized structures of all TMCs are provided in the ESI.
The geometry optimization as well as the calculation of MPA, NPA, ChelpG, Hirshfeld and CM5 charges were performed in gas phase using the GAUSSIAN 16 Rev A.03 suite of program\cite{g16}. Scalar-relativistic effects were taken into account for all calculations using the second-order Douglass-Kroll-Hess (DKH2) method.\cite{DOUGLAS197489,PhysRevA.33.3742} We compared the gas-phase and PCM-calculated PACs (computed at geometries re-optimized in PCM) and the results have shown that the solvent effect on the PACs is negligible, and henceforth we use gas-phase calculations for solute structure and PACs. The AIM charges were computed with the Multiwfn program\cite{JCC:JCC22885} using the wave function file (.wfx file) obtained from the DFT calculations. A high density grid is required for accurate numerical representation of the electron density to ensure convergence of the calculated AIM charges. In present work, this convergence was achieved at a grid spacing of 0.02 {\AA}. For the calculation of ChelpG charges, the vdW radii of 2.17 {\AA}, 2.02 {\AA} and 1.81 {\AA} were used for Ru$^{2+}$, Fe$^{2+}$ and Cu$^{+}$ metal ions in their ground states, respectively, which were taken from the literature.\cite{B701197K} In the excited states the above-mentioned vdW radii might no longer be adequate. Therefore, we also investigate the effect of different vdW radii of the metals on the ChelpG PACs and solvation structures. 
To obtain accurate ESP values, a high point density for the fitting procedure is necessary. Sigfridsson and Ryde\cite{Sigfridsson1998ComparisonMoments} have suggested to use at least 2000 ESP points per atom. In this work, the grid spacing was set to 0.15 {\AA} and employed ca. 4000 ESP points per atom in order to ensure that the charges are well-determined. The RESP charges were calculated using the two-stage RESP algorithm implemented in the Antechamber package\cite{WANG2006247,JCC:JCC20035} which is part of AmberTools. The default value of 0.1 \textit{e} was used for the \textit{b} term and the values of 0.0005 \textit{e} and 0.001 \textit{e} were set for the $\beta$ term for the first and second stage, respectively (see eqn (5)). We have performed two sets of RESP calculations. In the first set we have only restricted atoms with the same chemical environment to have the same partial charges, while in the second set we have used additionally the MPA charges as target charges for the metal atoms and fixed them during the fitting procedure (calculations tagged by RESP(MPA)).
The computed PACs of the four TMCs using different methods are reported in Figs S1-S4 of the ESI.  
                  
\subsubsection{Classical Molecular Dynamics Simulations}

All classical MD simulations were carried out with the Desmond software package\cite{4090217} at constant-temperature and volume (NVT). The DFT-optimized geometries of the selected TMCs were solvated in water (four-site TIP4P model)\cite{doi:10.1063/1.445869} for $\mathrm{[Ru(bpy)_3]^{2+}}$ and $\mathrm{[Fe(bpy)_3]^{2+}}$ and in ACN solvent for $\mathrm{[Fe(bmip)_2]^{2+}}$ and $\mathrm{[Cu(phen)_2]^{+}}$. The three-site model of Gu{\`{a}}rdia \emph{et al.}\cite{doi:10.1080/08927020108024509} was adopted for ACN. The selection of these solvents was made to match the experimental conditions of the time-resolved scattering and spectroscopic measurements performed on the investigated TMCs.\cite{Damrauer54,doi:10.1021/jp306917x,C3CC43833C,doi:10.1021/ja069300s} Chloride (Cl$^-$) counterions were added  for neutralization of the total charge. The standard OPLS 2005\cite{doi:10.1021/ja9621760} Lennard-Jones (LJ) parameters were used to model the nonbonded dispersion and exchange repulsion interactions between the atoms of the solute and the solvent. The ground- and excited-state PACs used in classical MD simulations are obtained from the DFT calculations described in the previous section and are kept fixed during the simulations. \par
The MD simulations were performed in a cubic box with size length of 35 {\AA} under periodic boundary conditions (PBCs). To speed up the calculation of forces, the multistep RESPA integrator\cite{doi:10.1021/j100078a035} was used, where the nonbonded-near and nonbonded-far (long-range electrostatic) interactions were updated every 1 fs and 3 fs, respectively. A distance cut-off of 9 {\AA} was applied to separate short- and long-range Coulombic interactions, for the latter, the particle mesh Ewald (PME) was used. For the equilibration of the system, the default protocol in Desmond was used, which consists of three stages: 1) A 100 ps constant-NVT simulation in Brownian regime at a temperature of 10 K and restraining solute heavy atoms with a force constant of 50 kcal mol$^{-1}$ {\AA}$^{-2}$. 2) A 12 ps constant-NVT simulation with the same temperature and restrains as stage 1. 3) A 24 ps constant-NVT simulation with the temperature increased to 300 K and no restraints. The Berendsen thermostat\cite{Berendsen1984MolecularBath} was applied in the equilibration. Finally, a 2 ns NVT production simulation was run by applying restraints on all solute atoms with a harmonic force constant of 1000 kcal mol$^{-1}$ {\AA}$^{-2}$ and the trajectory was recorded every 50 fs. The bond lengths involving hydrogen atoms in the solute were constrained using the M-SHAKE algorithm \cite{M-SHAKE} implemented in Desmond. The counterions were placed away from the solute and restrained with the same force constant to avoid any coordination with the solute. The system temperature was maintained at 300 K using the No\'se-Hoover thermostat\cite{Nose1984AMethods,Hoover1985CanonicalDistributions}. The structure and configuration input files were generated with the Maestro program (Schr\"{o}dinger, LLC). 

\begin{figure}[h]
    \centering
    \begin{subfigure}[b]{0.45\textwidth}
        \includegraphics[width=\textwidth]{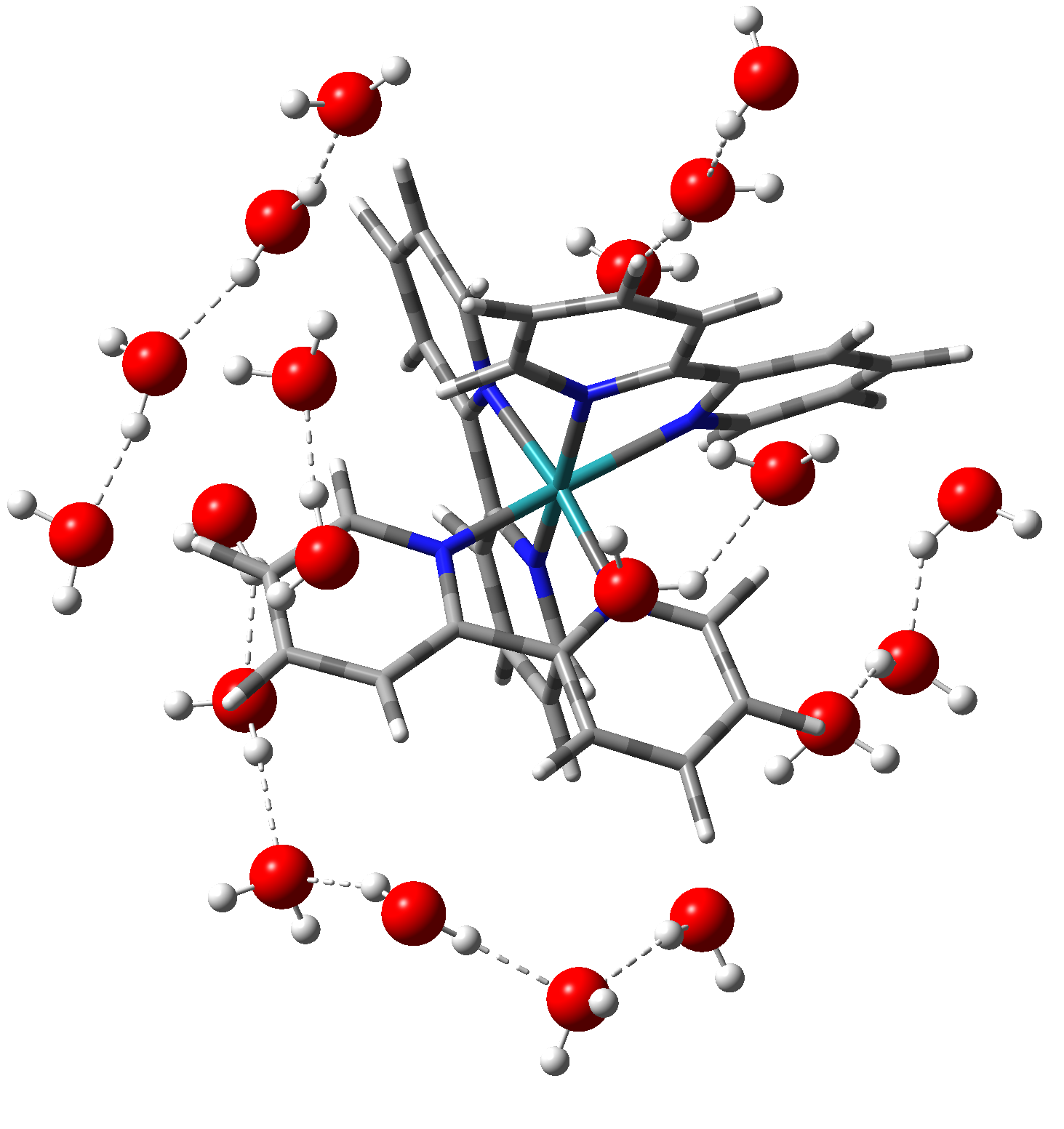}
    \end{subfigure}
   \caption{Graphical representation of the chain of hydrogen-bonded water molecules around $\mathrm{[Ru(bpy)_3]^{2+}}$ obtained from a snapshot of a QM/MM MD trajectory in the GS.}
\end{figure}
   
\subsubsection{QM/MM MD Simulations}
The QM/MM MD simulations were performed using the MD tools of the Atomic Simulation Environment (ASE) \cite{ASE2, ASE1} and the implementation of QM/MM electrostatic embedding \cite{Dohn2017,doi:10.1021/jz500850s} that interfaces ASE built-in classical force fields with the GPAW DFT code \cite{GPAW2, GPAW1}. For all four TMCs, the simulations employed a fixed QM/MM partitioning scheme, in which the complex (QM part) is entirely described with GPAW and the MM solvent is modeled through a fixed point-charge force field. The Kohn-Sham orbitals within the GPAW simulation cell for the QM solute were represented in a basis of linear combination of atomic orbitals (LCAO) \cite{GPAW_LCAO}, using TZP basis set \cite{GPAW_LCAO} for the metal and DZP basis set \cite{GPAW_LCAO} for the rest of the atoms. We assessed the performance of the selected mixed basis set against TZP basis set for all atoms in predicting the charge transfer in the MLCT state. The results show very similar charge transfer. The grid spacing of the cell was set to 0.18 {\AA}; this value was found to ensure convergence with respect to structural parameters of TMCs\cite{Levi2018SolutionSimulations}. Since forces for hybrid functionals are not yet implemented in GPAW, the BLYP functional, which is the GGA precursor of the hybrid functional B3LYP*, was used for describing all TMCs except $\mathrm{[Ru(bpy)_3]^{2+}}$ where the GGA DFT functional BP86 \cite{B88, P86} was applied. The BLYP has been used in previous studies.\cite{LawsonDaku2018Spin-stateStudy,doi:10.1021/jz100548m} The BP86 functional is known from previous DFT studies of this complex \cite{doi:10.1021/jp900147r, Buchs1998}, to give a GS structure and an energy separation between the lowest $^3$MLCT excited state and the GS in good agreement with the X-ray crystal structure and optical spectroscopic measurements, respectively.

In the following, we applied the same parameters as in the classical MD simulations described above, if not specified otherwise. 
The ACN force field was implemented in a development branch of ASE based on the parametrization of Ref. 67
and on the scheme for holonomic constraints of rigid triatomic molecules from Ref. 62. For the nonbonded interactions, a standard LJ potential was used, in which LJ parameters for the atoms of the complex were taken from the universal force field (UFF) \cite{UFF}. 
The QM/MM MD data in a solvent bath at 300 K were obtained for the GS of all four TMCs and for the excited state of $\mathrm{[Ru(bpy)_3]^{2+}}$, $\mathrm{[Fe(bpy)_3]^{2+}}$ and $\mathrm{[Cu(phen)_2]^{+}}$. The procedure that we employed for each of the four systems is the following. First, the GS geometry of the complex was optimized with GPAW in vacuum using a quasi-Newton local optimization algorithm implemented in ASE. Then, the GS optimized geometry was centered in a box of solvent molecules pre-equilibrated in the NVT ensemble at 300 K.
After solvating the complex, the QM/MM simulation box was equilibrated in the NVT ensemble to 300 K employing a time step of 1 fs. 
The Equilibration was carried out with the ASE Langevin thermostat applied to the atoms of the solvent. PBCs were treated according to the minimum image convention.\cite{Allen:1989}\par
During these simulations the solute geometry is flexible and in order to eliminate the fastest vibrational motions and thus reduce the computational time in the QM/MM MD simulations, we enforced two bond length constraints per hydrogen atom in the complex using the RATTLE algorithm \cite{RATTLE} as implemented in ASE. Following thermal equilibration of the solvent, QM/MM MD data were collected for at least 18 ps with a time step of 2 fs. From this first equilibrated trajectory, a set of other 20-45 QM/MM trajectories were started to accelerate the data collection. The starting MD frames were spaced by at least 0.5 ps from each other. Moreover, to further minimize the correlation between them, the velocities of the atoms of each of the starting frames were randomized by imposing a Maxwell-Boltzmann distribution at 300 K. Overall, we collected between 150 and 400 ps of 300 K equilibrated QM/MM MD data for the GS of each of the four complexes.\par
For $\mathrm{[Ru(bpy)_3]^{2+}}$, $\mathrm{[Fe(bpy)_3]^{2+}}$ and $\mathrm{[Cu(phen)_2]^{+}}$, we further generated QM/MM MD data in the same excited states as considered in the classical MD investigation. 
This was achieved by starting excited-state QM/MM trajectories from a set of representative configurations of each of the equilibrated GS trajectories. The excited states were described using a recent implementation of $\Delta$SCF in GPAW \cite{Levi2018SolutionSimulations}, based on fixing the electronic configuration of the system with Gaussian smeared constraints on the orbital occupation numbers. The Gaussian smearing ensures stable convergence of the electronic density at each step during the QM/MM MD propagation. We used a flexible width for the Gaussian functions controlling the extent of the smearing during the SCF cycle. Starting from an initial value of 0.01 \AA{}, the width was increased by 0.01 \AA{} at each 120 SCF steps until convergence of the density. In most of the cases, convergence of the SCF cycle took place within the first 120 steps. The $\Delta$SCF-QM/MM trajectories were propagated with a time step of 2 fs, with the Langevin thermostat applied to the solvent. In total, we collected between 100 and 200 ps of excited-state $\Delta$SCF-QM/MM trajectories for each of the three TMCs. In the cases of $\mathrm{[Fe(bpy)_3]^{2+}}$, we observe that the solvation shell and solute structure relax within 3 ps. In $\mathrm{[Cu(phen)_2]^{+}}$, as copper is tetracoordinated, the planes of the two ligands are perpendicular in the GS and due to a pseudo Jahn-Teller distortion, flat in the $^3$MLCT state. This flattening in our QM/MM MD simulations occurs within 3 ps, which is in a good accordance with the experimental record in the $^1$MLCT state\cite{Iwamura2015UltrafastComplexes}. The average atomic Cartesian coordinates and significant internal structural parameters of the solvated TMCs obtained as averages from the QM/MM MD trajectories are reported in Tables S1-14 of the ESI. For a comparison, structural information from the gas-phase DFT optimizations using the Gaussian 16 and GPAW programs are also reported.\par
In terms of computational efficiency, our simulations indicate that the QM/MM MD simulations, using 16 CPU cores, are 4 orders of magnitude slower than the classical MD simulations.

\subsubsection{Analysis of Solvation Structure}

Solute-solvent RDFs, \textit {g(r)}, from the ground- and excited-state classical MD and QM/MM MD simulations were computed using the VMD software\cite{HUMPHREY199633} with a bin size of  0.1 {\AA} for the radial sampling. For the excited state simulations, we ensured that the RDFs reflected equilibrium distributions by checking the convergence with respect to the amount of sampled configurations included in the computation of the RDFs. Furthermore, the running solvent coordination number ($cn$), as shown in eqn (6), was used to obtain information about the solvent organization and orientation around the complex.
\begin{equation}
cn(R) = 4\pi\rho \int_{0}^{R} r^2 g_{\mathrm{m-s}}(r) dr 
\end{equation}

Here, $\rho$ is the density of the bulk solvent. $cn(R)$ gives the number of s solvent atoms in a sphere with radius $R$ around the transition metal center m.

\begin{figure}[h]
    \centering
    \begin{subfigure}[b]{0.45\textwidth}
        \includegraphics[width=\textwidth]{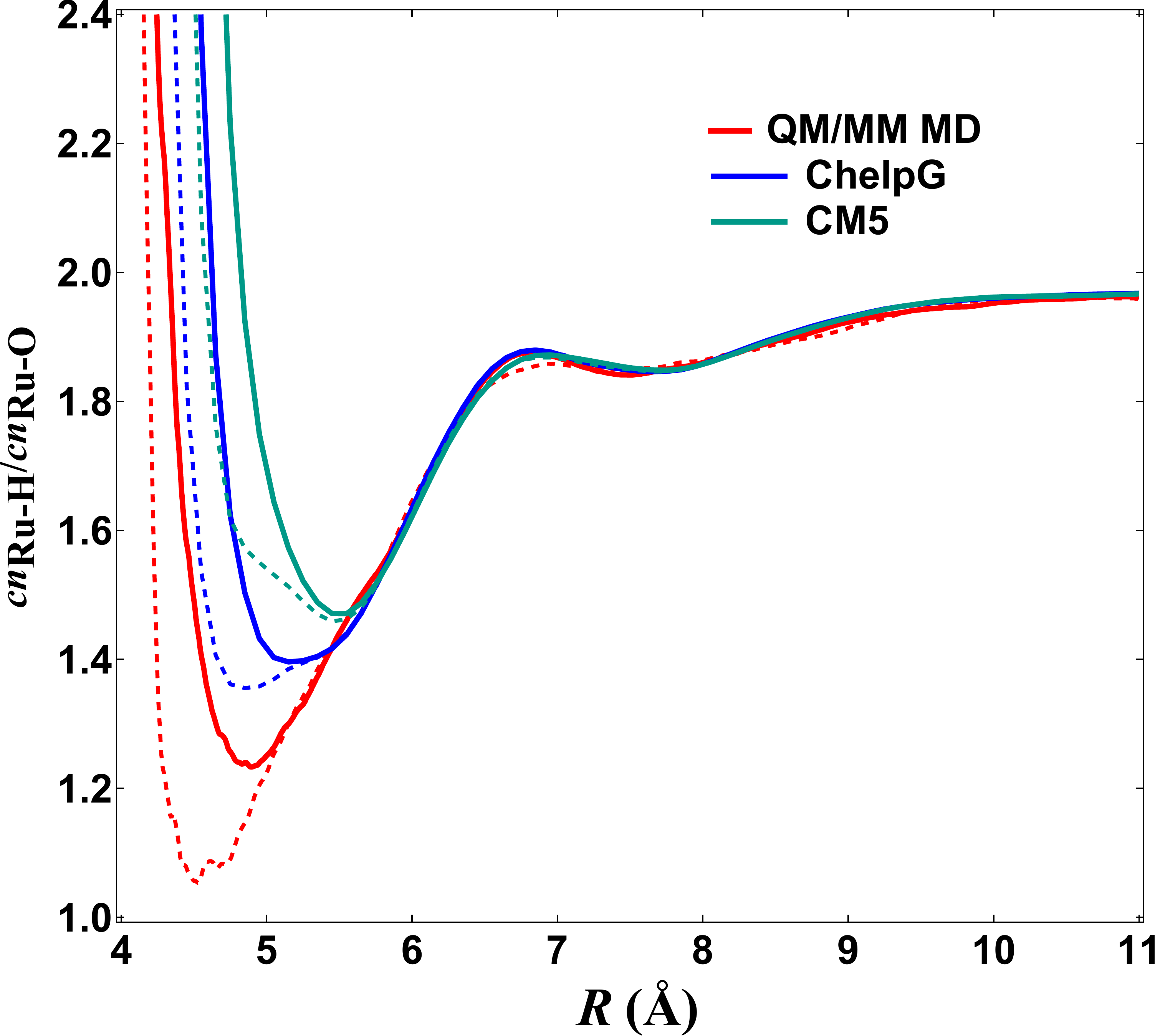}
    \end{subfigure}
   \caption{Plot of the coordination number ratio $\mathrm{{\textit{cn}}_{Ru-H{_w}}/{\textit{cn}}_{Ru-O{_w}}}$ as a function of the distance from the Ru atom for $\mathrm{[Ru(bpy)_3]^{2+}}$ in water obtained from the classical MD simulations, using the ChelpG and CM5 methods, and QM/MM MD simulations in the GS (solid lines) and $^3$MLCT state (dashed lines).}
\end{figure}

\begin{figure*}[h]
    \centering
    \begin{subfigure}[b]{0.8\textwidth}
        \includegraphics[width=\textwidth]{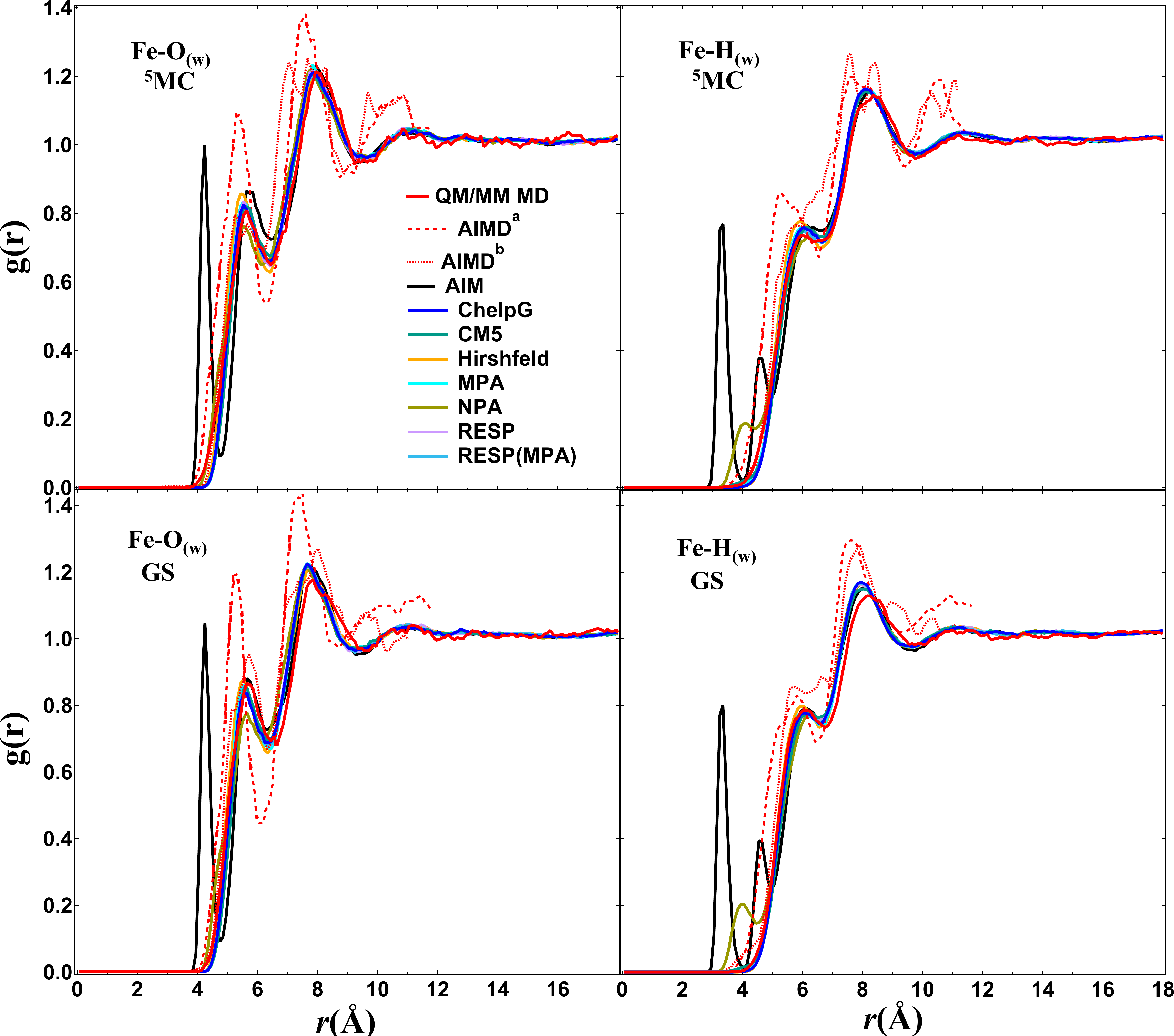}
    \end{subfigure}
    \caption{ The RDFs, $g(r)$, of $\mathrm{[Fe(bpy)_3]^{2+}}$ in water for the Fe-$\mathrm{O_w}$ and Fe-$\mathrm{H_w}$ pairs obtained from the classical MD and QM/MM MD (solid red lines) simulations in the GS and $^5$MC state. It is also shown the AIMD data from the literature: $^a$ Refs. \citenum{LawsonDaku2018Spin-stateStudy} and $^b$ Refs.\citenum{doi:10.1021/jz100548m}. The AIMD data (dashed and dotted lines) are available until 12 {\AA}.}
\end{figure*}

\section{Results}
Fig. 2 shows the RDFs of $\mathrm{[Ru(bpy)_3]^{2+}}$ in water for the $\mathrm{Ru-O_w}$ and $\mathrm{Ru-H_w}$ pairs, which are labeled by $\mathrm{{\textit{g}}_{Ru-O_w}}(r)$ and $\mathrm{{\textit{g}}_{Ru-H_w}}(r)$. The first peak of the $\mathrm{g_{Ru-O_w}}(r)$ bears important information about the first solvation shell. As seen in Fig. 2, left panels, for the results of QM/MM MD in either GS or $^3$MLCT state, this peak is located at 5.5 {\AA} and is followed by a valley at 6.45 {\AA}. 
The $\mathrm{{\textit{cn}}_{Ru-O{_w}}}$ shows that this shell carries an approximate number of 15 water molecules for both the GS and $^3$MLCT state
and contains a chain of hydrogen-bonded water molecules intercalated between the bpy ligands (Fig. 3). Moret \textit{et al.} \cite{doi:10.1021/jp900147r,CHEM:CHEM201000184} using MD simulations in the GS and QM/MM MD simulations in the $^3$MLCT state, and also Tavernelli \textit{et al.}\cite{TAVERNELLI2011101} for the $^1$MLCT state by QM/MM MD simulations, observed the same solvation structure. The red lines in Fig. 4 show plots of the coordination number ratio $\mathrm{{\textit{n}}_{HO}}(r)$ = $\mathrm{{\textit{cn}}_{Ru-H{_w}}({\textit{r}})/{\textit{cn}}_{Ru-O{_w}}}(r)$ obtained from the QM/MM MD simulations in the GS (solid line) and the $^3$MLCT state (dashed line). At short distances from the Ru center (below 4 {\AA}, not shown in Fig. 4), very large $\mathrm{{\textit{n}}_{HO}}$ values reveal that only water H atoms can approach the Ru atom. The $\mathrm{{\textit{n}}_{HO}}$ falls down below 2 between 4.3 {\AA} and 6.6 {\AA}, which indicates that the oxygen atoms of water orient toward Ru in the first solvation shell. 
Thereafter, $\mathrm{{\textit{n}}_{HO}}$ then converges toward 2 reflecting the random orientation of water in the bulk solvent. Having fairly different $\mathrm{n_{HO}}$ for the GS and the excited state, despite negligible changes in the corresponding RDFs (Fig. 2), reflects the high sensitivity of this parameter to the small changes in solvent organization. By comparing the minima of the ratios in the case of the QM/MM MD, it is found that upon transition from the GS to the $^3$MLCT state, water molecules prefer to re-orient through the oxygen atoms toward the Ru$^{2+}$ cation at \textasciitilde 0.35 {\AA} shorter distance. 
  
The results obtained from the classical MD simulations show that the RDFs from the ChelpG, CM5, Hirshfeld, MPA, RESP and RESP(MPA) methods reproduce the QM/MM MD RDFs very well. The NPA method, except small shoulders at \textasciitilde 4 {\AA} in $\mathrm{{\textit{g}}_{Ru-H_w}}(r)$, also predicts the RDFs in good agreement with the QM/MM MD ones but AIM fails. 
Fig. S1 shows that the AIM method predicts large positive charge (+1.2 \textit{e}) for the Ru and large negative charges (-1.1 \textit{e}) for the nitrogen atoms bonded to Ru, which causes larger charge separations in the bpy ligands, compared to the other PAC methods (see Fig. S6). The large negative charges and more accessibility of the nitrogen atoms with respect to Ru, provide a condition for intercalation of three water molecules between the bpy ligands and hydrogen bonding with the hydrogen atoms of water. 

For the ChalpG charges in the $^3$MLCT state, we studied the effect of different vdW radii for Ru on the charges and RDFs. The vdW radii of 2.17 {\AA}, 1.80 {\AA} and 1.20 {\AA} have been used for the Ru atom in the calculation of ChelpG charges and their corresponding RDFs are shown in Fig. S7. The results show that although different vdW radii can affect the PACs significantly, this has no considerable effect on the RDFs. Fig. 4 and S8 present a comparison between the $\mathrm{n_{HO}}$ obtained from the QM/MM MD and classical MD simulations. The results show that the ChelpG, RESP, RESP(MPA) and Hirshfeld methods successfully reproduce the $\mathrm{n_{HO}}$ ratio of QM/MM MD while the AIM and NPA and MPA methods fail. 
Upon going from the GS to the $^3$MLCT state, the differences between RDFs are very small indicating that the amount of charge transfer from the Ru to the bpy ligands is not sufficient to change the equilibrium solvation structure in the $^3$MLCT state. Fig. S6 supports this observation by showing the same charge separations of the bpy ligands for the GS and the $^3$MLCT state. 

The charge localization in the excited state of $\mathrm{[Ru(bpy)_3]^{2+}}$ in water from the QM/MM MD simulations has also been investigated. The DFT gas-phase calculations give charge distribution and, thus, PACs distributed equally over the three bpy ligands. Fig. S5 illustrates spin densities calculated from different snapshots along a single QM/MM MD trajectory in the 3MLCT state. During the first ps, the charge oscillates between all ligands and then it localizes over at most two bpy ligands, with the pair carrying the charge changing during the dynamics. These results are in good agreement with the observation of Moret \textit{et al.}\cite{CHEM:CHEM201000184}. We note that using fixed equally distributed charges in classical MD simulations does not seem to affect the RDFs compared to the QM/MM MD ones. This can be attributed to the small magnitude of the charge transfer upon excitation (0.1-0.3 $e$, as obtained from different PAC methods), which can also be seen from very similar RDFs for the ground and excited states.

\begin{figure}[h]
    \centering
    \begin{subfigure}[b]{0.45\textwidth}
        \includegraphics[width=\textwidth]{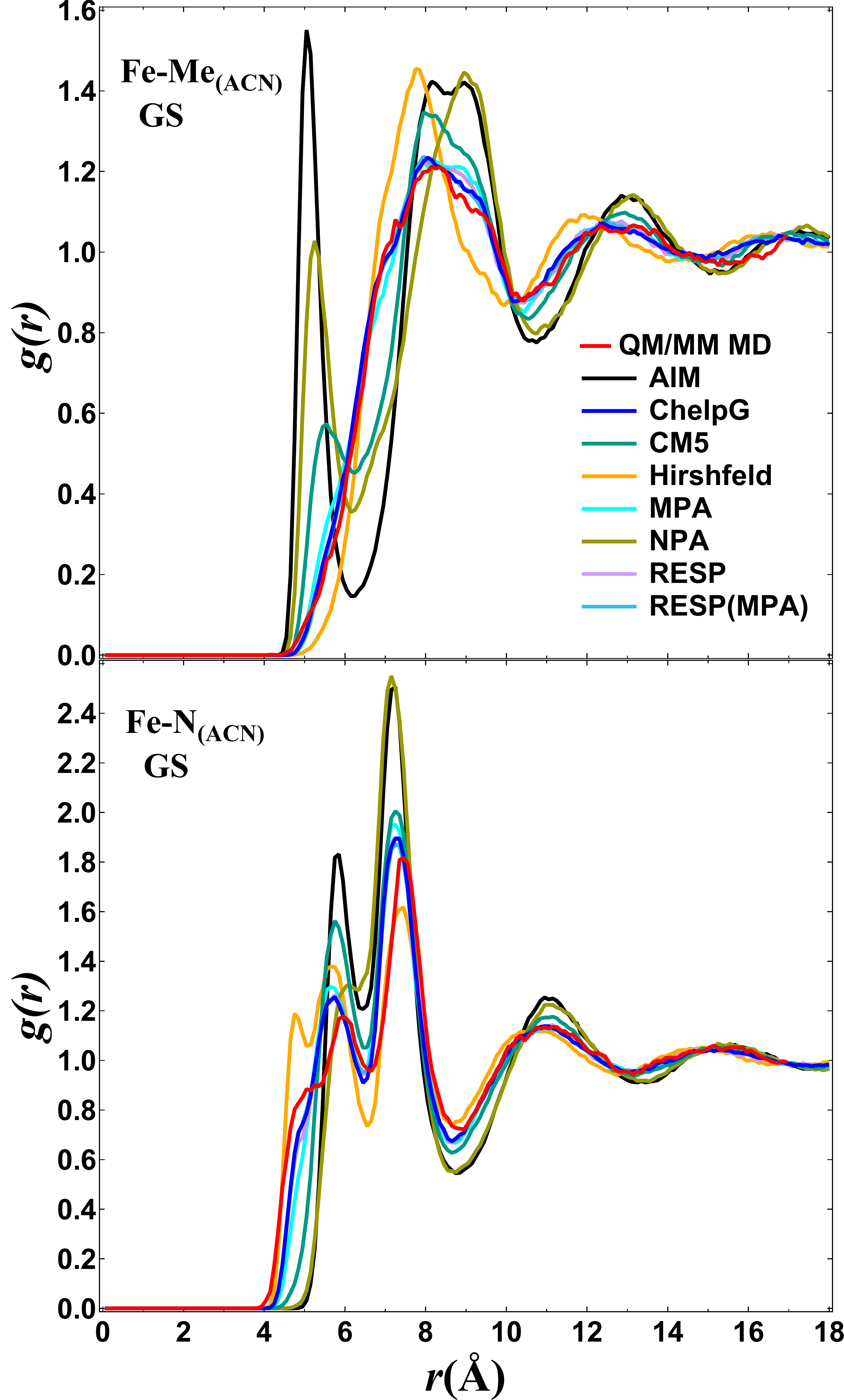}
    \end{subfigure}
      \caption{The RDFs, $g(r)$, of $\mathrm{[Fe(bmip)_2]^{2+}}$ in ACN for the Fe-$\mathrm{N_{(ACN)}}$ and Fe-$\mathrm{Me_{(ACN)}}$ pairs obtained from the classical MD and QM/MM MD (red lines) simulations in the GS. Note that some curves are overlapped with each other.}
\end{figure} 
   
\begin{figure*}[h]
    \centering
   \begin{subfigure}[b]{0.8\textwidth}
       \includegraphics[width=\textwidth]{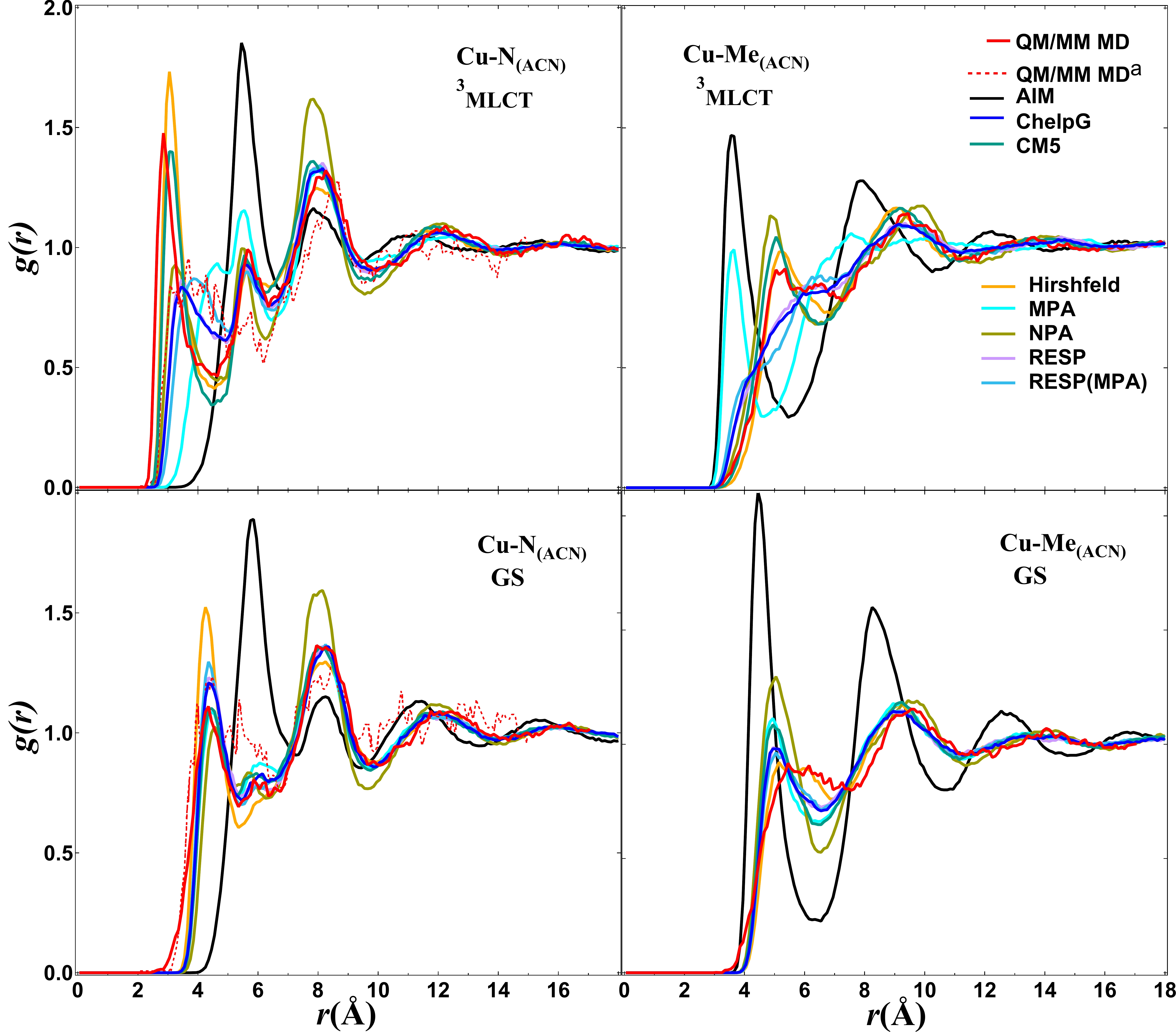}
    \end{subfigure}
    \caption{The RDFs, $g(r)$, of $\mathrm{[Cu(phen)_2]^{+}}$ in ACN for the Cu-$\mathrm{N_{(ACN)}}$ and Cu-$\mathrm{Me_{(ACN)}}$ pairs obtained from the classical MD and QM/MM MD (red lines) simulations in the GS and $^3$MLCT state. Also shown are the QM/MM MD data from the literature: $^a$Ref. 19.}
\end{figure*}

\begin{figure*}[h]
    \centering
   \begin{subfigure}[b]{0.8\textwidth}
       \includegraphics[width=\textwidth]{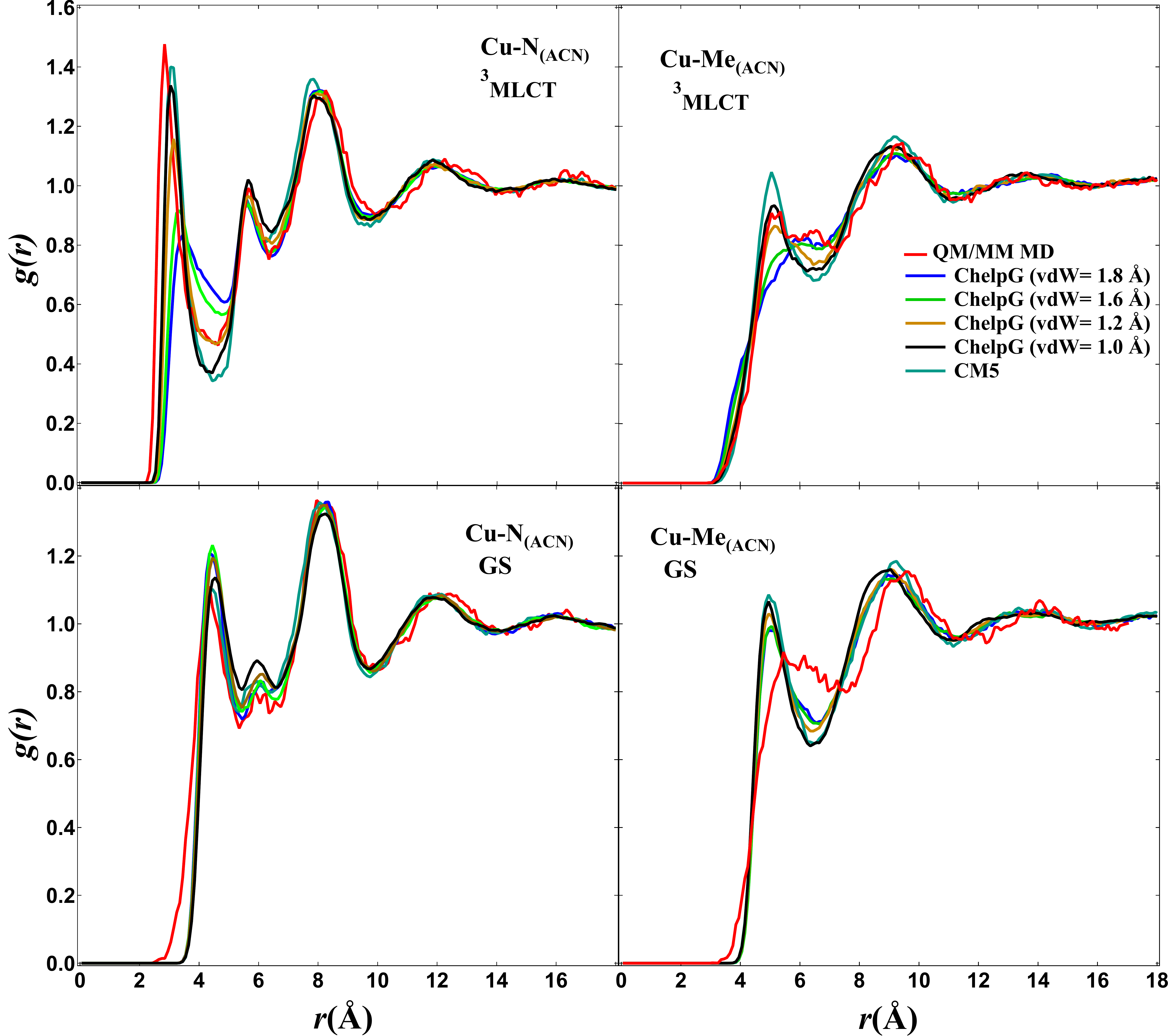}
    \end{subfigure}
   \caption{The RDFs, $g(r)$, of $\mathrm{[Cu(phen)_2]^{+}}$ in ACN for the Cu-$\mathrm{N_{(ACN)}}$ and Cu-$\mathrm{Me_{(ACN)}}$ pairs obtained from the classical MD, using the ChelpG method with different vdW radii of Cu, and QM/MM MD (red lines) simulations in the GS and $^3$MLCT state.}
\end{figure*}

Fig. 5 illustrates the $\mathrm{{\textit{g}}_{Fe-O_w}}(r)$ and $\mathrm{{\textit{g}}_{Fe-H_w}}(r)$ of $\mathrm{[Fe(bpy)_3]^{2+}}$ extracted from the QM/MM MD and classical MD simulations in water. As expected from the similarity between the ground state ligand structure of $\mathrm{[Ru(bpy)_3]^{2+}}$ and $\mathrm{[Fe(bpy)_3]^{2+}}$ and type of solvent, the same trends in the RDFs of these two complexes are observed. 
Lawson Daku \textit{et al.}\cite{LawsonDaku2018Spin-stateStudy,doi:10.1021/jz100548m} have investigated $\mathrm{[Fe(bpy)_3]^{2+}}$  utilizing AIMD simulations. The red dotted and dashed lines in Fig. 5 are the AIMD RDFs. The dotted one\cite{doi:10.1021/jz100548m} was obtained using the BLYP functional within the Car-Parinello MD (CPMD) scheme and the simulations were performed for 24.5 and 4 ps for the GS and the $^5$MC state, respectively. Their results have shown that upon going from the GS to the $^5$MC states, two water molecules are expelled from the first solvation shell ($\sim$17 in the GS and $\sim$15 in the $^5$MC). The dashed lines\cite{LawsonDaku2018Spin-stateStudy} correspond to RDFs obtained by applying the dispersion-corrected BLYP-D3 functional in the BOMD approach, in order to describe long-range dispersion interactions, and calculated for longer simulation times, 76.6 ps for the GS and 67.2 ps for the $^5$MC state. The new study\cite{LawsonDaku2018Spin-stateStudy} revealed that the number of water molecules in the first shell actually increases from $\sim$15 in the GS state to $\sim$17 in the $^5$MC state. The RDFs and the resulting $cn_{\mathrm{Fe-O_W}}$ value obtained from our QM/MM MD, by going from the GS to the $^5$MC state, show that around 0.7 water molecule is expelled from the first coordination shell into the bulk solvent. 
The expulsion of water molecules from the first solvation shell is consistent with the increase in the density of bulk solvent by 0.2\% upon formation of the $^5$MC state, as measured by Haldrup \textit{et al.} using XDS.\cite{Haldrup2016ObservingScattering} The RDFs extracted from classical MD simulations for all PAC methods, except AIM, are in good agreement with the QM/MM MD and show the same trend: a decrease in the number of water molecules in the first shell upon the GS $\rightarrow$ $^5$MC transition. We investigated the effect of changing the vdW radii of Fe on the excited-state ChelpG charges and RDFs. As for $\mathrm{[Ru(bpy)_3]^{2+}}$, we did not see any effect on the RDFs by using vdW radii of 2.02 {\AA} and 1.2 \AA. (see Fig. S9 in ESI) 

From the $\mathrm{[Ru(bpy)_3]^{2+}}$ and $\mathrm{[Fe(bpy)_3]^{2+}}$ RDF results, it can be concluded that the structure of the three bidentate bpy ligands prevent the solvent molecules to coordinate \textit{directly} to the metals and be affected by their charges. This shows that the ligand charges have a more important role than the metal charges in the determination of the solvation structure in such TMCs. However, the charges of the metals have an \textit{indirect} effect on the solvation structure by changing the charges of the neighboring nitrogen atoms. 

Fig. 6 shows the $\mathrm{{\textit{g}}_{Fe-N_{(ACN)}}}(r)$ and $\mathrm{{\textit{g}}_{Fe-Me_{(ACN)}}}(r)$ for the GS of $\mathrm{[Fe(bmip)_2]^{2+}}$ in ACN. As mentioned before (Section 2.2.1), the existence of several conical intersections between the low-lying triplet MLCT and MC excited states did not allow us to carry out state-specific QM/MM MD simulations for the $^3$MLCT state of $\mathrm{[Fe(bmip)_2]^{2+}}$. $\mathrm{[Fe(bmip)_2]^{2+}}$ has two tridentate bmip ligands and gives the possibility to the solvent molecules (ACN) to approach the metal atom in the simulations. This is reflected in the larger differences of the RDFs in $\mathrm{[Fe(bmip)_2]^{2+}}$ compared to the two previous cases. The AIM and NPA methods, similarly to the cases of $\mathrm{[Ru(bpy)_3]^{2+}}$ and $\mathrm{[Fe(bpy)_3]^{2+}}$, exaggerate the negative charges of the nitrogen atoms (see Fig. S3), leading to attraction of the methyl groups of the ACN molecules. These electrostatic attractive interactions are reflected in the structured peaks centered at 5 {\AA} and 5.2 {\AA} in the $\mathrm{{\textit{g}}_{Fe-Me_{(ACN)}}}(r)$ corresponding to the AIM and NPA methods, respectively (see Fig. 6, top panel). 
However, here the peaks are located at longer distances. This is due to the bulky structure of the ACN molecules, compared to water, which prevents them from intercalating between the ligands and getting close to Fe. 
Among the applied PAC methods, the MPA, ChelpG, RESP and RESP(MPA) methods provide RDFs close to the QM/MM MD ones. Although, we do not have QM/MM MD results for the $^3$MLCT state of $\mathrm{[Fe(bmip)_2]^{2+}}$ to compare with, for the reason explained above, we have, for the sake of completeness, included the $\mathrm{{\textit{g}}_{Fe-N_{(ACN)}}}(r)$ and $\mathrm{{\textit{g}}_{Fe-Me_{(ACN)}}}(r)$ obtained from classical MD simulations in the ESI (see Fig. S10).\par
The last case that we have considered for this study is $\mathrm{[Cu(phen)_2]^{+}}$ in ACN. The main reason for choosing this TMC is its unique ligand structure that offers the possibility of direct coordination of solvent molecules to the copper. 
This enables us to study the \textit{direct} effect of the metal charge on the calculated RDFs using different PAC methods. Fig. 7 displays the $\mathrm{{\textit{g}}_{Cu-N_{(ACN)}}}(r)$ and $\mathrm{{\textit{g}}_{Cu-Me_{(ACN)}}}(r)$ for $\mathrm{[Cu(phen)_2]^{+}}$ in the GS and $^3$MLCT state in ACN. The red dashed lines in Fig. 7 show QM/MM MD RDFs, which were taken from Ref. 19 and were calculated from $\sim$20 ps simulations using the CPMD scheme. 
Owing to the flattened geometry of $\mathrm{[Cu(phen)_2]^{+}}$ in the $^3$MLCT state, we are able to assess the performance of each PAC method more precisely. By inspection of the coordination number of the first solvation shell in the QM/MM MD $\mathrm{{\textit{g}}_{Cu-N_{(ACN)}}}(r)$ it is realized that upon transition from the GS to the $^3$MLCT state this shell shifts to a shorter distance by 1.5 {\AA}. This is showing an increased Cu-N coordination and, at the same time, a decrease in the number of ACN molecules from 4.2 in the GS to 2.6 in the $^3$MLCT state, i.e., a shift of $\sim$1.5 ACN molecules to the second shell upon transition to the $^3$MLCT state.
Among the applied PAC methods in classical MD simulations, for the GS all of them, except AIM and NPA, reproduce the QM/MM MD RDFs reasonably well. On the other hand, in the $^3$MLCT state, only CM5 along with Hirshfeld, which predict large positive charges on Cu (see Fig. S4), provide RDFs relatively close to the QM/MM MD ones. Note that although AIM and NPA provide large positive charges on Cu, the negative charges on the nitrogen atoms cause repulsive forces between the complex and the ACN molecules. In case of the GS $\mathrm{{\textit{g}}_{Cu-N_{(ACN)}}}(r)$, none of the PAC methods can reproduce the tail in the QM/MM MD RDF at short distance (r<3 \AA). This may be attributed to the flexibility of the solute not included in the classical MD simulations (see Fig. S16).

Similar to $\mathrm{[Ru(bpy)_3]^{2+}}$, we study the effect of the vdW radius of Cu on the excited-state ChelpG PACs and RDFs. The values of 1.80 {\AA}, 1.60 {\AA}, 1.20 {\AA} and 1.00 {\AA} were used in computing the ChelpG charges. The calculated PACs reveal that by decreasing the vdW radius, the charges of Cu and N atoms become more positive and negative, respectively, i.e., the ionic character is increased. As seen in Fig. 8, top panels, the RDF results in the $^3$MLCT state show that using different vdW radii in ChelpG PACs calculations leads to remarkable changes in the RDFs. This is ascribed to the significant effect of the charges of Cu and N atoms on the calculated RDFs of $\mathrm{[Cu(phen)_2]^{+}}$. However, according to Fig. 8, bottom panels, the GS RDFs are much less sensitive to the chosen vdW radius. The results show that applying vdW radii of 1.20 {\AA} or 1.00 {\AA} for Cu in the excited-state ChelpG calculations can provide RDFs fairly close to the CM5 and QM/MM MD. 

\section{Discussion and Conclusions}
In this work, utilizing the RDF, which is a powerful tool for characterizing the solvation structure, we have evaluated the
performance of several most-used PAC methods in classical MD simulations aimed at describing ground- and excited-state solvation structures. Several PAC methods have been considered for this study including MPA, NPA, ChelpG, RESP, RESP(MPA), AIM, Hirshfeld and CM5. For this purpose, four popular polypyridine TMCs have been chosen: $\mathrm{[Ru(bpy)_3]^{2+}}$, $\mathrm{[Fe(bpy)_3]^{2+}}$, $\mathrm{[Fe(bmip)_2]^{2+}}$ and $\mathrm{[Cu(phen)_2]^{+}}$. We analyse the RDFs obtained from classical MD simulations using fixed charges and frozen solute structure, and compare them to more accurate QM/MM MD RDFs where both the electronic and nuclear structures are allowed to evolve. These results show that for the four investigated TMCs, the AIM and NPA methods are not suitable to characterize the solvation structure of TMCs that possess ionic character. This is not surprising because these methods suffer from overestimation of PACs for atoms in ionic bonds.49,50 Depending on the ligand structure, the ChelpG, RESP, RESP(MPA) and CM5 methods are well-suited to describe the solvation structure. For $\mathrm{[Ru(bpy)_3]^{2+}}$ and $\mathrm{[Fe(bpy)_3]^{2+}}$, the three bidentate bpy ligands do not allow the solvent molecules to feel the charge of the metal directly. 
For such TMCs like $\mathrm{[Ru(bpy)_3]^{2+}}$ and $\mathrm{[Fe(bpy)_3]^{2+}}$, with exclusion of AIM, one can apply any other PAC method of this work in classical MD simulations and obtain RDFs as accurate as those provided by QM/MM BOMD in water. In $\mathrm{[Fe(bmip)_2]^{2+}}$ and especially $\mathrm{[Cu(phen)_2]^{+}}$, more space is available between the ligands which enables the solvent molecules to approach the metals. In such cases, our results in ACN indicate that a careful selection of the PAC method is required. Thus, the selection of PAC method is dependent on the coordination number of the metal and the denticity of the ligands. 

To extend our conclusion, we also studied the effect of the type of the solvent. To do so, we have repeated the classical MD simulations for $\mathrm{[Ru(bpy)_3]^{2+}}$ and $\mathrm{[Fe(bpy)_3]^{2+}}$ in ACN solvent and for $\mathrm{[Fe(bmip)_2]^{2+}}$ and $\mathrm{[Cu(phen)_2]^{+}}$ in water; Figs. S11-S14 present their corresponding RDFs. The results show that the RDFs simulated in ACN are more sensitive to the choice of the applied PAC methods than in water, particularly in the cases of $\mathrm{[Fe(bmip)_2]^{2+}}$ and $\mathrm{[Cu(phen)_2]^{+}}$ for which more space is accessible between the ligands. 
These results might be due to the nearly twice as large dipole moment of ACN (3.96 D) compared to the one of water (2.18 D), as computed from the applied force fields\cite{doi:10.1080/08927020108024509,doi:10.1063/1.481505}.
Hence, in addition to the ligand denticity and metal coordination number dependency, the selection of PAC method is also solvent-dependent. The effect of the vdW radius of the metals on ChelpG PACs and resulting RDFs for the GS and $^3$MLCT state of $\mathrm{[Ru(bpy)_3]^{2+}}$ and $\mathrm{[Cu(phen)_2]^{+}}$ in ACN was also studied. By decreasing this parameter, the charge of the metal becomes more positive while the bonded N atoms get more negative and leading to a higher ionic character. For $\mathrm{[Ru(bpy)_3]^{2+}}$, no changes are observed in the RDFs obtained using different vdW radii. However, the excited state of $\mathrm{[Cu(phen)_2]^{+}}$ is found to be highly sensitive on the vdW radius. Our calculations show that by using vdW radii of 1.20 or 1.00 {\AA} for the Cu atom in the ChelpG calculations, we can produce RDFs in good agreement with the QM/MM MD.
Furthermore, for the $^3$MLCT state of $\mathrm{[Cu(phen)_2]^{+}}$, we studied this effect in water. These results show that the obtained RDFs (see Fig. S15) are more sensitive to changing the vdW radius in ACN than water. This leads to the conclusion that the application of ChelpG PACs for such cases requires the optimization of the vdW radius of the metal by further benchmarking.\\
According to the RDF results, only PAC methods derived from physical observables, the ESP for the ChelpG, RESP, and RESP(MPA) methods and the molecular dipole moment for the CM5 method, enable us to produce RDFs as close as those from QM/MM MD. As mentioned in Section 2.1, the RESP method is designed to overcome the problem of charge prediction of deeply buried atoms in the ChelpG method. However, in all four cases in this work, the results have demonstrated almost identical RDFs for the ChelpG and the RESP/RESP(MPA) methods. 
Among the applied PAC methods, the ChelpG/RESP and the CM5 PAC methods can characterize the solvation structure around TMCs using fast classical MD simulations with the accuracy of QM/MM MD simulations.\\ 
 
{\Large\textbf{Conflicts of interest}}\\
There are no conflicts to declare.\\

{\Large\textbf{Acknowledgements}}
The research leading to the presented results has received funding from the People Programme (Marie Curie Actions) of the European Union's Seventh Framework Programme (FP7/
2007-2013) under REA Grant Agreement No. 609405 (COFUNDPostdocDTU), the Danish Council of Independent Research Grant No. 4002-00272, the Independent Research Fund
Denmark Grant No. 8021-00347B, and was also supported by the "Lend\"{u}let" (Momentum) Program of the Hungarian Academy of Sciences (LP2013-59). GL thanks the Icelandic Research Fund for financial support.

\balance


\bibliography{Mendeley} 
\bibliographystyle{rsc} 

\end{document}